# Unique features of polarization in ferroelectric ionic conductors

Andrew O'Hara[1], Nina Balke[2], and Sokrates T. Pantelides[1,3*]

[1]*Department of Physics and Astronomy, Vanderbilt university, Nashville, TN 37235*
[2]*Center for Nanophase Materials Sciences, Oak Ridge National Laboratory, Oak Ridge, TN 37831*
[3]*Department of Electrical and Computer Engineering, Vanderbilt University, Nashville, TN 37235*

## Abstract

Ferroelectrics that are also ionic conductors offer possibilities for novel applications with high tunability, especially if the same atomic species causes both phenomena. In particular, at temperatures just below the Curie temperature, polarized states may be sustainable as the mobile species is driven in a controlled way over the energy barrier that governs ionic conduction, resulting in unique control of the polarization. This possibility was recently demonstrated in $CuInP_2S_6$, a layered ferroelectric ionic conductor in which Cu ions cause both ferroelectricity and ionic conduction. Here, we show that the commonly used approach to calculate the polarization of evolving atomic configurations in ferroelectrics using the modern theory of polarization, namely concerted (synchronous) migration of the displacing ions, is not well suited to describe the polarization evolution as the Cu ions cross the van der Waals gaps. We introduce an asynchronous Cu-migration scheme, which reflects the physical process by which Cu ions migrate, resolves the difficulties, and describes the polarization evolution both for normal ferroelectric switching and for transitions across the van der Waals gaps, providing a single framework to discuss ferroelectric ionic conductors.

*pantelides@vanderbilt.edu

## Introduction

Ferroelectrics are materials that, below the Curie temperature, feature spontaneous distortions that generate a remanent polarization *P*, analog of the remanent magnetization of ferromagnetic materials. Common ferroelectric crystalline materials such as perovskite-structure transition-metal oxides, some phosphates, carbides, nitrides, etc., typically have two stable, oppositely oriented polarization states described by a double-well potential along one or more crystallographic directions. The spontaneous atomic displacements from the centrosymmetric configuration are typically small, of order 0.1 or 0.2 Å.[1] The two energy minima, separated by an energy barrier $\Delta E_1$, represent two equivalent atomic configurations at temperatures below the Curie temperature. Applying an external electric field $\mathcal{E}$ and sequentially reversing it can drive the displacing atoms back and forth between the two energy minima, generating hysteresis *P-$\mathcal{E}$* loops. In experimental constructions of such loops, polarization is typically obtained by integrating the current, measuring the charge on a reference capacitor in series with the sample,[2] or by piezoelectric force microscopy (PFM), where one measures the piezoelectric coefficient and relates it to the polarization.[3]

According to the modern theory of polarization (MTP),[4-7] the polarization ***P*** of a periodic solid is defined only modulo "a quantum of polarization" that reflects the lattice periodicity. As a result, only polarization differences $\Delta P$ and derivatives of polarization, for example the piezoelectric coefficient $d_{ijk} = dP_i/dT_{jk}$, where $T_{jk}$ is the stress tensor, can be measured directly. Indeed, many applications rely on the evolution of polarization under the influence of an applied electric field. Similarly, one typically measures switching between polarization states and constructs *P-$\mathcal{E}$* loops. MTP calculations of the evolution of polarization are typically performed using density-functional theory (DFT) by assuming that the sheets of displacing ions move in a synchronous (concerted) way, whereby one easily constructs a function $P(z)$.[1]

These polarization values are "referenced" to the zero-polarization centrosymmetric configuration, i.e., they represent $\Delta P$ values that can be measured.[4-7]

In ferroelectrics that are also ionic conductors, the description of polarization is a more complex process if the same species mediates both properties, as first discussed by Scott.[8] In such materials, adjacent double-well potentials are linked by an energy barrier $\Delta E_2$, which governs the activation energy of ionic conductivity. Depending on the relative magnitudes of the two energy barriers, it may be possible under controlled conditions to drive the displacing ions over $\Delta E_2$ to the next polarization state and perhaps beyond, before disordered ionic conduction becomes dominant. This possibility was recently demonstrated in $CuInP_2S_6$ (CIPS), a layered van-der-Waals (vdW) ferrielectric ionic conductor with Cu ions causing both phenomena (indium atoms undergo relatively small counter displacements so that CIPS is often referred to simply as ferroelectric).[3,9] CIPS is recently attracting significant interest, in part because of its unusual polarization properties.[3,9-15]

Calculations of the total energy $\Delta E(z)$ as well as $P(z)$ using the synchronous Cu-migration (SCM) approach in the layer regions of CIPS were recently reported in Ref. [12]. They revealed the existence of a quadruple well, repeated periodically in each layer, describing a total of two low-polarization (±LP) and two high-polarization (±HP) states (see the blue box in Figure 1A). The extension of $P(z)$ across the vdW gaps reported in Ref. [3], however, entails an error, which led to the conclusion that the SCM is entirely suitable for CIPS and can account for the experimental data (an Erratum has been submitted[16]). Here, we revisit the SCM approach and demonstrate that, in CIPS, it is intrinsically limited in its ability to describe fully the processes that occur when the Cu ions cross the vdW gaps. As a more suitable alternative, we introduce an asynchronous Cu migration (ACM) approach that is motivated by fundamental reasoning as follows (we discuss CIPS for concreteness, but the approach should be suitable for all ferroelectrics that are also ionic conductors with the same atomic species causing both phenomena):

1) We first examine the rationale that makes the SCM approach suitable for displacive ferroelectrics, e.g., $BaTiO_3$, in which the displacements are small, of order 0.1-0.2 Å.[1] In such materials, the centrosymmetric atomic configuration with the sheets of displacing ions at the mid-point between the two potential wells is the stable zero-polarization paraelectric phase above the Curie temperature. Below the Curie temperature, this is unstable, i.e., it has an imaginary-frequency polar phonon mode that induces the spontaneous displacements. In such materials, ionic conduction by the species that cause polarization is generally not possible. Thus, it is appropriate to adopt this centrosymmetric configuration as the reference state in MTP calculations of $P(z)$ and move the displacing atoms in a synchronous way.

2) CIPS, however, is an order-disorder ferroelectric with very large Cu displacements from the nominal mid-layer centrosymmetry plane (1.57 Å and 2.14 Å, for the LP and HP states, respectively; in the HP states, the Cu sheets are just outside the layers in the vdW gaps and actually bond to S atoms across the gaps). Furthermore, the paraelectric phase of CIPS above the Curie temperature is a statistical average of individual Cu ions occupying random polarization states, adding up to a net zero polarization[17,18] (analogous to the Ising model of paramagnetic states in magnetism). In fact, the evolution of the ferroelectric to the paraelectric phase as the Curie temperature is approached has been mapped out by experiments showing snapshots of a gradual redistribution of the Cu ions from all the Cu atoms being in the LP ground state to all the available energy minima with equal fractions in the pairs of positive and negative polarization states (see Figure 2 of Ref. [17])

3) Quantum molecular dynamics simulations have shown that Cu ions make transitions over the mid-layer or the mid-gap energy barriers independently, one at a time, at picosecond time scales, whereas full transitions between adjacent polarization states have been observed over times of order a few to 100 milliseconds.[3,9]

Given the above, we define an ACM process for a transition from polarization state $P_1$ to the adjacent polarization state $P_2$ via individual Cu-ion jumps and envision a series of snapshots at intermediate times when distinct fractions $f$ of Cu atoms, e.g., 10%, 20%, 30%, and so on, have transitioned to $P_2$. For each fraction $f$, the average position of the transitioning Cu atoms defines a $z$ value, which happens to be the same as the position of *hypothetical* Cu sheets in the SCM approach. For MTP calculations, there is a unique zero-polarization centrosymmetric configuration in which 50% of the Cu atoms occupy each of the ±LP or the ±HP sites, akin to the experimental observations of Ref. [17] that we discussed earlier. In the remainder of the paper, we examine in detail the polarization curves $P(z)$ for synchronous and asynchronous Cu migration. The two curves are identical within the layers but are markedly different in the vdW gap regions, reflecting the fact that the motion of the Cu ions in the vdW gaps does not affect the polarization evolution the same way that it does within the layers. We demonstrate that, unlike the SCM scheme, the ACM scheme provides a consistent framework to account for the available experimental data. First, therefore, we summarize the pertinent experimental data for CIPS.

Macroscopic $P$-$\mathcal{E}$ loops constructed by integrating the switching current have so far been reported only for the LP state within the CIPS layers.[11,13,17,19] However, all four predicted polarization states in CIPS were detected at length scales of order tens to hundreds of nm using PFM measurements of the longitudinal piezoelectric coefficient $d = d_{333}$ (the energy barrier between the LP and HP states is only a few meV, but local strains stabilize one or the other of the two states).[12] In the laboratory reference system, the measured $d$ is related to the polarization $P$ by[20-22]

$$d = 2QP\varepsilon\varepsilon_0. \qquad (1)$$

Here $Q$ is the effective fourth-order electrostriction-coefficient tensor, $\varepsilon$ is the dielectric tensor, and $\varepsilon_0$ is the vacuum permittivity [in Eq. (1), the symbols of vectors and tensors refer to their longitudinal components in the $z$ direction]. Since $Q$ is a property of the material that does not depend on the direction of $P$, the sign of the measured $d$ value dictates the sign of the corresponding $P$ value. In Ref. [12], good agreement between the measured $d$ values and the DFT-predicted values referenced to the zero-polarization centrosymmetric state enabled the assignment of the four DFT-calculated polarization values to the four measured $d$ values.

Subsequent PFM experiments using a capacitor geometry[9] employed a series of electric-field pulses of successively longer duration and measured the $d$ values of the polarization state at the end of each pulse within the layers and across the vdW gaps. The polarization states were identified by relating the measured and calculated $d/P$ ratios. A key point was the fact that, upon transitioning from the +HP state across the vdW gaps to the -HP state, the measured $d$ value simply changes sign. Since no phase transition is involved, Eq. (1) signals that *the polarization aligns against the external electric field*.[9] In a different PFM experiment[3] using voltage spectroscopy[23] on bare surfaces, $d$-$\mathcal{E}$ loops were measured and converted to $P$-$\mathcal{E}$ loops using the $d \leftrightarrow P$ correspondence.[12] These loops confirmed the observation of the transition from the +HP state across the vdW gaps to the -HP state and found that this transition exhibits $dP/d\mathcal{E} < 0$, which signals that the region features a novel form of negative capacitance.[3]

## 2. The synchronous Cu-migration (SCM) approach

In this Section, we report DFT calculations of $P(z)$ using the SCM approach as the Cu sheets are moved through the layers and across the vdW gaps, left to right in Figure 1A, and explore the applicability of the approach to explain available experimental data. In the SCM approach, there exist two $z$ positions for the Cu sheets, the mid-layer and mid-gap planes, that result in centrosymmetric atomic configurations.

Thus, it is equally legitimate to calculate $P(z)$ using either of these configurations as the reference state. Using DFT and the MTP, we obtain the blue and red curves shown in Figures 1A and 1B for the two reference states, respectively. The intrinsic periodicity of the system leads to identical $P(z)$ curves that are shifted up or down by one or more quanta (the dashed blue curves in Figure 1A and the dashed red curves in Figure 1B; a quantum in CIPS is ~50 µC/cm$^2$). Furthermore, the blue and red curves have identical shapes: a red curve matches the corresponding blue curve if it is shifted up or down by a half a quantum and vice versa. As a result, the polarization of one centrosymmetric configuration is not zero when referenced to the other such configuration; instead, it has a value of half a quantum.

In Figure 1A, the blue rectangle highlights the quadruple well as described in Ref. [12], with the LP values in round numbers being ±4 µC/cm$^2$ and the HP values being ±11 µC/cm$^2$, referenced to the centrosymmetric atomic configuration with the Cu sheets at the mid-layer planes.[3] In Figure 1B, the red rectangle highlights a different, but equally legitimate, perspective of the same quadruple well, with the polarization curves referenced to the centrosymmetric atomic configuration with the Cu sheets at the mid-gap planes. The polarization values assigned to the four polarization states now exhibit both a sign and magnitude-ordering reversal. Retaining the original labels of the four polarization states (see Figures 1A and 1B), we have the reassignments LP: ±4 → ∓21 µC/cm$^2$ and HP: ±11 → ∓14 µC/cm$^2$. The SCM approach *predicts* that, in CIPS, one can, in principle, obtain standard $P$-$\mathcal{E}$ loops with different remanent polarization values by polarization switching through the layers and by polarization switching through the vdW gaps. In particular, for HP loops in the layers, the remanent polarization values are ±11 µC/cm$^2$ ($\Delta P = 22$ µC/cm$^2$), whereas in the vdW gaps the remanent polarization values are ±14 µC/cm$^2$ ($\Delta P = 28$ µC/cm$^2$).

The only $P$-$\mathcal{E}$ loops across the vdW gaps, from +HP to -HP, that have been reported so far were obtained by measuring $d$-$\mathcal{E}$ loops and converting them to $P$-$\mathcal{E}$ loops.[3] These loops are inverted relative to standard $P$-$\mathcal{E}$ loops and the remanent polarizations are ±11 µC/cm$^2$ ($\Delta P = -22$ µC/cm$^2$). The origin of the discrepancy between the SCM predictions and the experimental $d$-$\mathcal{E}$ and $P$-$\mathcal{E}$ loops is the fact that in the SCM scheme the polarization *rises* as the Cu sheets cross the vdW gaps, according to the curves in Figure 1. On the other hand, the PFM-measured $d$ values, measured before the voltage is turned off, retain their absolute magnitude and change their sign when all the Cu ions arrive at the -HP sites. By Eq. (1), therefore, the polarization value also changes sign, from +11 to -11 µC/cm$^2$, i.e., the polarization *dips* by -22 µC/cm$^2$ and aligns against the electric field, confirming the findings of Ref. [9]. The net conclusion here is that the SCM prediction is not suited to describe the polarization changes when the Cu ions transition across the vdW gaps because the SCM scheme does not reflect the physical reality of how Cu ions migrate. In the next section, we show that the $P(z)$ curve calculated using the ACM approach is in full agreement with the PFM data and provides fundamental justification for the $d \leftrightarrow P$ association. We will also discuss additional insights into why the SCM works in describing polarization transitions within the layers, but fails in the vdW gap regions.

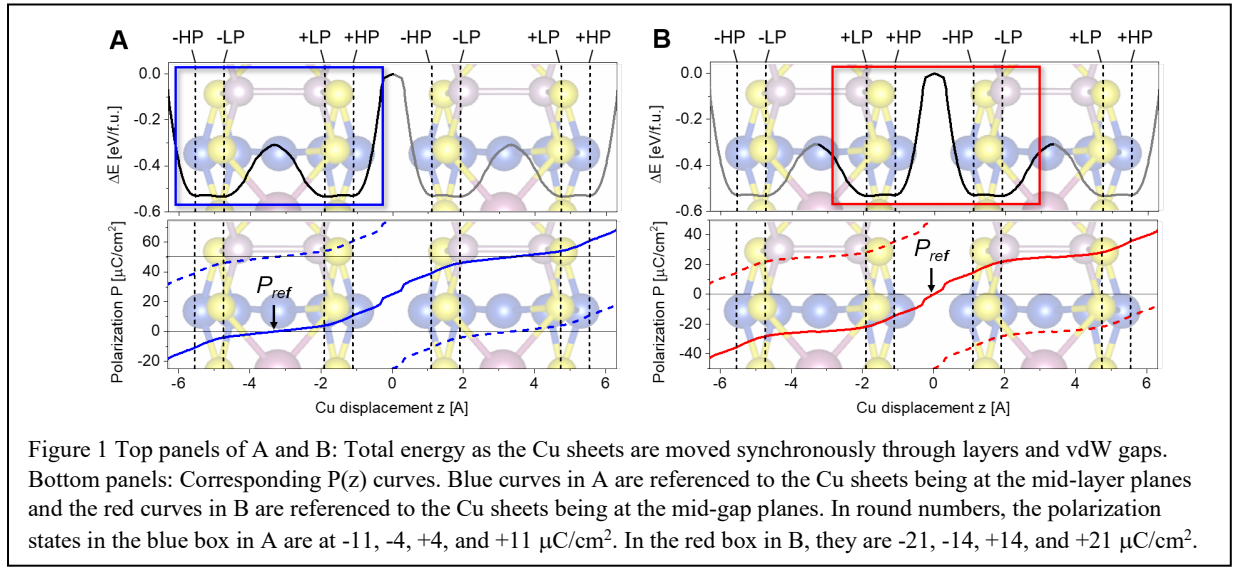

Figure 1 Top panels of A and B: Total energy as the Cu sheets are moved synchronously through layers and vdW gaps. Bottom panels: Corresponding P(z) curves. Blue curves in A are referenced to the Cu sheets being at the mid-layer planes and the red curves in B are referenced to the Cu sheets being at the mid-gap planes. In round numbers, the polarization states in the blue box in A are at -11, -4, +4, and +11 µC/cm$^2$. In the red box in B, they are -21, -14, +14, and +21 µC/cm$^2$.

## 3. The asynchronous Cu-migration (ACM) approach

The limitations of the SCM approach that we identified above are caused by the fact that, by the very definition of SCM, as the Cu sheets migrate in a concerted way, the polarization increases monotonically. As we explained in the introduction, however, individual Cu atoms migrate independently from a polarization state $P_1$ to the next polarization $P_2$ on picosecond time scales, but we can monitor at intermediate time scales the arrival at $P_2$ of fractions, e.g., 10%, 20%, etc. of the total Cu atoms in the system. We can then compute the polarization of the evolving structure. In this section, we show that the ACM approach, which, unlike SCM, reflects the physical reality of Cu migration, can lead to either a rise or a drop in $P(z)$ and accounts for the experimental data we discussed so far.

In constructing ACM $P(z)$ curves, we assume that a transition from polarization state $P_1$ to polarization state $P_2$ is completed before Cu atoms start "leaking" to the next polarization state $P_3$. This assumption is consistent with the assumption made in the SCM scheme, namely the Cu sheets arrive at state $P_2$ before any Cu migration to state $P_3$ commences. This idealized assumption leads to cusps in $P(z)$ (see Figure 2 below). Inclusion of "leaks" to the next polarization state and other effects (thermal vibrations and the electric field pushing the Cu ions slightly to the right of the nominal polarization-state sites) would round off the cusps.

The simplest implementation of the ACM scheme can be achieved without any step-by-step calculations by simply adopting the values of the four polarization states in the quadruple well, i.e., all Cu atoms reside in one of the four quadruple-well minima. We recall that the DFT-MTP results are: LP = ±4 µC/cm² and ±HP = ±11 µC/cm², as shown in the blue box in Figure 1A, referenced to the zero-polarization state in which all Cu atoms reside in the mid-layer plane. The same values are valid when referenced to the ACM 50-50 zero-polarization state. To calculate a $P(z)$ curve, we envision Cu atoms driven to the right by an electric field, but, as usual, the effect of the electric field is not explicitly included. At each successive intermediate step during the transition from polarization state $P_1$ to polarization state $P_2$, a growing fraction $f$ of Cu atoms has already transitioned to the $P_2$ state while a fraction $(1-f)$ of Cu atoms is still at $P_1$. We can then sum the two contributions and get the approximate result

$$P(z) = (1-f)P_1 + fP_2, \qquad (2)$$

where $z$ is the average position of the transitioning Cu atoms. For a sequence of transitions between adjacent polarization states $P_1$ and $P_2$ ($P_{1/2}$ = -LP/+LP, +LP/+HP, +HP/-HP, -HP/-LP, etc.), the result amounts to a piecewise straight line as shown in Figure 2.

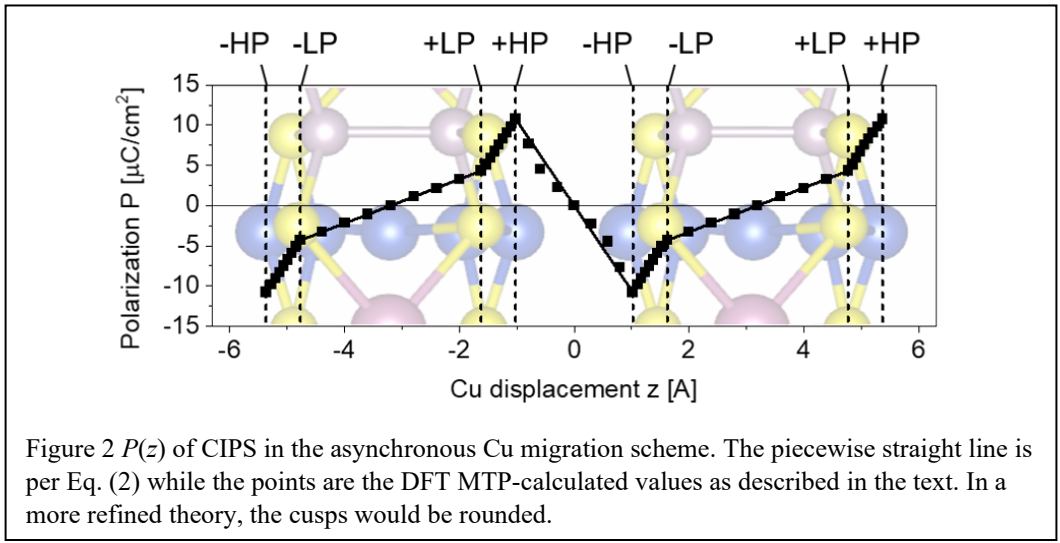

Figure 2 $P(z)$ of CIPS in the asynchronous Cu migration scheme. The piecewise straight line is per Eq. (2) while the points are the DFT MTP-calculated values as described in the text. In a more refined theory, the cusps would be rounded.

We now turn to DFT calculations of the ACM $P(z)$. We recall that, in the ACM scheme, we effectively have a single zero-polarization reference state, namely the paraelectric configuration of Cu atoms randomly occupying one of the four polarization states with a net $P = 0$. At $T = 0$ K where we carry out DFT-MTP calculations, this state can be realized as 50% Cu in each of the ±LP states (the ground state of the system). The choice 50% Cu in each of the ±HP states or a mixed ±LP/±HP state adding up to zero polarization are equally valid and give identical $P(z)$ curves because, unlike the case in the SCM scheme, the different realizations of the asynchronous reference state retain their zero value no matter which of them is used as reference. As a result, we get a unique *microscopic* $P(z)$ curve that reflects the polarization values of the underlying atomic configurations, as deduced from PFM measurements of the piezoelectric coefficient $d$ and calculated $d \leftrightarrow P$ correspondence.

We employed DFT as in Ref. [3] to perform MTP calculations of $P(z)$ for the same sequence of $P_{1/2}$ transitions, referenced to the zero-polarization 50-50 arrangement of Cu atoms in ±LP or ±HP. In each $P_{1/2}$ case, the MTP calculation was performed for a series of fractions $f$ as follows. We used a supercell containing eight Cu atoms per layer so that transitioning one of these Cu atom amounts to a snapshot in which a fraction $f = 1/8$ (12.5%) of all Cu atoms have transitioned from $P_1$ to $P_2$ (recall that Cu atoms transition independently over picosecond time scales while $f = 1/8$ is one of eight snapshots as the Cu atoms accumulate in $P_2$; we found that the MTP-calculated $P$ values vary only slightly for different possible configurations of Cu atoms at each fraction $f$; we found that such averaging has a negligible effect). Both layers in the supercell are assumed to be stoichiometric in each snapshot (i.e., a layer always has 8 Cu atoms). The LP lattice constant was used for transitions between LP states across the layer, the HP lattice constant was used for transitions between HP states across the vdW gap, and an interpolated lattice constant was used for LP to HP transitions. The calculated $P(z)$ curve is shown in Figure 2, where we see that it compares very well with the approximate piecewise straight line.

In order to illustrate the ACM process, in Figure 3 we show schematics of the calculated relaxed atomic configurations in both the SCM and ACM schemes at different stages of the Cu migration through the different polarization states in the layers and across the vdW gaps. In some of the ACM panels, some of the Cu atoms are not fully visible as they are obstructed by sulfur atoms. The blue bars underneath each schematic designate what fraction of the layer's Cu atoms reside where in the system. The small black tic marks designate the LP- and HP-polarization sites, which straddle the layer edges. The differences in the two schemes are quite vivid: In panel (a), we see Cu sheets at the mid-layer planes in synchronous scheme, while in the asynchronous schemes the Cu atoms are 50-50 at the ±LP sites of each layer. In panel (b) the synchronous Cu sheets have moved half-way toward the +LP site, while in the asynchronous scheme 75% of the Cu atoms in each layer remain at the -LP sites and 35% of them have moved to the +LP sites. In panel (c), both the synchronous and asynchronous Cu atoms reside at the +LP sites. Panels (d) show the Cu atoms at the +HP sites, again the same in the two schemes. In panels (e), the synchronous Cu sheets have moved from the +HP sites a quarter of the distance to the -HP sites across the vdW gaps and in panel (f) to the mid-gap plane. The contrast with the respective asynchronous configurations is evident, as 25% and then 50% of all Cu atoms move from the +HP sites to the -HP sites instead.

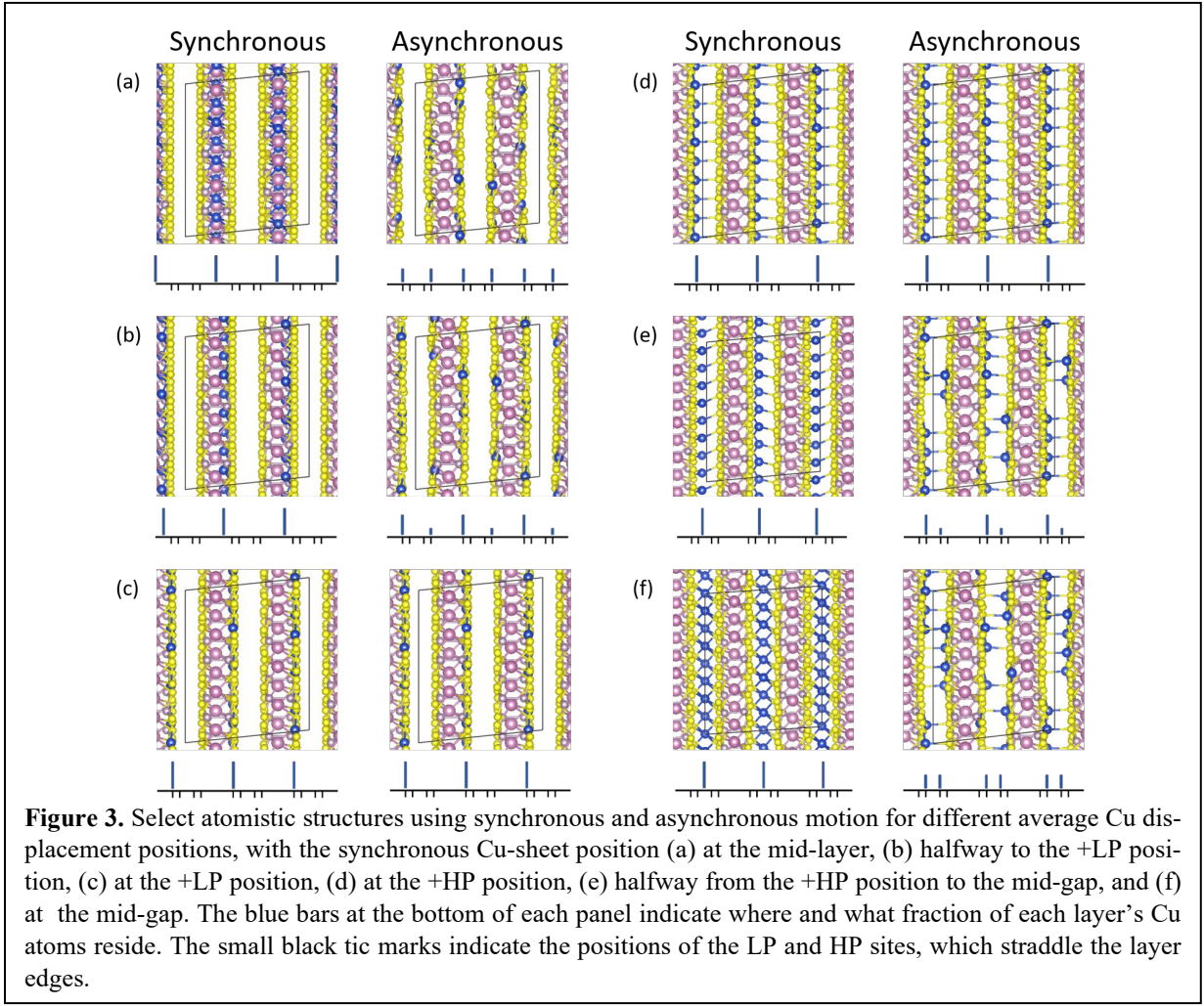

**Figure 3.** Select atomistic structures using synchronous and asynchronous motion for different average Cu displacement positions, with the synchronous Cu-sheet position (a) at the mid-layer, (b) halfway to the +LP position, (c) at the +LP position, (d) at the +HP position, (e) halfway from the +HP position to the mid-gap, and (f) at the mid-gap. The blue bars at the bottom of each panel indicate where and what fraction of each layer's Cu atoms reside. The small black tic marks indicate the positions of the LP and HP sites, which straddle the layer edges.

A comparison of Figure 2 with Figure 1 reveals the following. The ACM $P(z)$ curve in the layer regions in Figure 2 is identical to the corresponding SCM blue curve in Figure 1A. On the other hand, the ACM and SCM curves differ markedly in the vdW-gap regions. When the Cu atoms cross the vdW gaps, the ACM curve *predicts* that the system transitions from a standard +HP atomic configuration (referenced to a zero-polarization 50-50 Cu atoms in ±HP states) with $P = +11$ µC/cm$^2$ to a standard -HP atomic configuration referenced to the same zero-polarization state with $P = -11$ µC/cm$^2$ and $\Delta P = -22$ µC/cm$^2$, which the SCM scheme could not capture. Thus, the ACM-calculated microscopic $P(z)$ is fully consistent with measurements of piezoelectric coefficients of polarization states that have been accessed by applying electric pulses as in Refs. [3,9] The ACM curve of Figure 2 confirms that the polarization is indeed aligned against the electric field in these experiments.

A notable feature in the asynchronous $P(z)$ curve of Figure 2 is that $dP/dz > 0$ for all $P_{1/2}$ transitions in the layer regions where all transitions are to states of higher polarization; however, $dP/dz < 0$ in the vdW-gap regions, where the $P_{1/2}$ transition is to a lower-polarization state. One normally identifies $dP/dz$ with the effective charge of the mobile species upon which the electric field acts. In the present case, however, we know that, at the atomic scale, the Cu ions are transitioning from left to right independently, one at a time, at picosecond time scales, as in the quantum MD simulations of Refs. [9,12]. Here, d$P$/d$z$ refers to an abstract effective charge, which in the gap regions moves backwards and results in a polarization decrease, as we shall now describe.

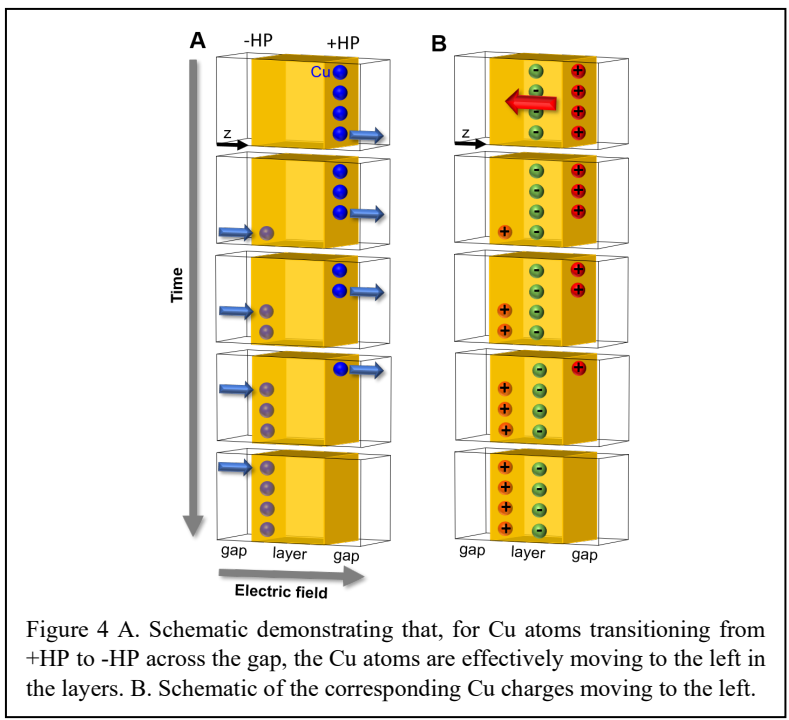

Figure 4 A. Schematic demonstrating that, for Cu atoms transitioning from +HP to -HP across the gap, the Cu atoms are effectively moving to the left in the layers. B. Schematic of the corresponding Cu charges moving to the left.

In order to identify and explain the origin of the microscopic currents that underlie the negative slope of the ACM $P(z)$ in the vdW-gap regions, we turn to Figure 4. In Figure 4, we depict the evolution of one layer, drawn as an orange box, with a vdW gap on either side, and show a schematic of how the Cu atoms move asynchronously from left to right. We use only four Cu atoms (shown as blue balls) per layer in the $xy$ supercell for illustration purposes, so that each blue dot corresponds to 25% of all Cu atoms in the system. Initially, the four Cu atoms are on the right surface of the orange box (the +HP site). As time progresses from top to bottom, we monitor snapshots of the Cu atoms transitioning at 25%, 50%, 75% and 100% to the -HP state to the right into the next layer, which is not shown (every time a blue dot transfers from the orange box to the right, a bluish-gray ball, also representing 25% of Cu atoms, enters the box from the left and sits on the left surface of the orange box (the -HP site). The sequence of schematics in Figure 4A are used to calculate the polarization values with DFT and the MTP (the calculations whose results are plotted in Figure 2 are actually based on $xy$-supercells containing eight Cu atoms per layer as discussed above). These polarization values are placed in Figure 2 at the $z$ values that correspond to the average position of the Cu atoms that are transitioning across the vdW gaps. It is clear by the arrows in the sequence of snapshots in Figure 4A that the Cu atoms are physically moving to the right, with their average position within the vdW gaps. However, *within each layer*, the Cu atoms appear to be moving from right to left. As a result, the apparent currents from right to left control the evolution of $P(z)$ across the vdW gaps, causing the polarization values of the snapshots to decrease as the Cu atoms transition asynchronously from +HP to -HP. This polarization decrease is also evident in Figure 4B, where we show the evolution of the nominal charges in the sequence of snapshots that are used to perform the DFT MTP calculations. The positive Cu charges appear to straddle the mid-layer plane *from right to left* as the negative charges of the InP$_2$S$_6^-$ sheets remain in place at the mid-layer plane (the In atoms actually move slightly in the counter direction). It is the atomic configurations that correspond to these nominal charges that generate the MTP-calculated $P(z)$ values in Figure 2 with d$P$/d$z$ < 0. As discussed in Ref. [3], when the $P$-$\mathcal{E}$ loop across the vdW gaps is generated from the $d$-$\mathcal{E}$ loop, one finds $dP/d\mathcal{E} < 0$, which signals a transient negative capacitance of a different origin than the commonly discussed mechanism.[24-26] Thus, the $P(z)$ curve in Figure 2 is in full accord with the experimental data in Refs. [3,9].

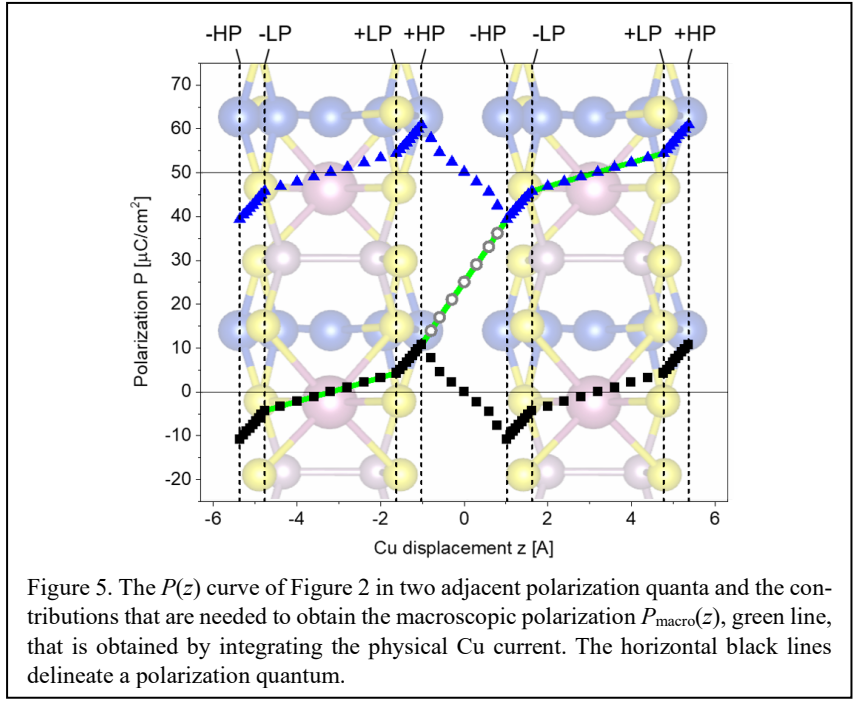

Figure 5. The $P(z)$ curve of Figure 2 in two adjacent polarization quanta and the contributions that are needed to obtain the macroscopic polarization $P_{macro}(z)$, green line, that is obtained by integrating the physical Cu current. The horizontal black lines delineate a polarization quantum.

It is, in fact, intriguing to compare the asynchronous Cu migration scheme with the construction of the $d$-$\mathcal{E}$ loop in Ref. [3], which is then converted to a $P$-$\mathcal{E}$ loop by associating $d$ values with $P$ values as discussed earlier. The reported hysteresis curves are based on measurements of $d$ that detect the area fraction of +HP and -HP polarization states as function of electric field pulse, which correspond to our snapshots of $f$ fractions that have transitioned from the +HP to the -HP sites across the vdW gaps. We conclude that the ACM scheme accounts naturally for all the polarization results obtained by converting PFM-measured $d$ values to $P$ values.[3,9]

Finally, we wish to carry out a direct comparison of the SCM and ACM $P(z)$ curves. In the SCM scheme, in the presence of an electric field, each Cu atom is migrating with the same velocity, which then automatically becomes the effective "drift" velocity of the ionic component of the total macroscopic current. The DFT-calculated $P(z)$ then corresponds to the integral of this total current. In a gedanken experiment, if we could apply a sequence of identical electrical pulses that are designed to just move the Cu sheets from the -HP sites to the next -HP sites, the integral of the total current, which we would normally call the "ionic current", would simply go up by a polarization quantum at a time, crossing consecutive quanta according to the blue curves in Figure 1A. In order to illustrate the power of the ACM $P(z)$ curve of Figure 2, we show that we can recover the SCM $P(z)$, which corresponds to the integral of the total current, as follows.

The schematic of Figure 4A tells us that the ACM microscopic-polarization values in the gap regions in Figure 2 arise from Cu ions that appear to move from right to left. Thus, to recover the SCM macroscopic polarization that corresponds to integrating the total SCM current, we need to add a contribution from $Cu^+$ ions moving to the right by a distance $R$ between the -HP site of the greyish-blue balls in Figure 4A to the -HP sites of the next layers where the blue balls of Figure 4A transit (not shown). This distance is the same $R$ that enters the definition of a polarization "quantum" $P_Q$, namely $P_Q = 2eR/\Omega$.[4-7] A simple calculation then gives

$$P_{macro}(z) = P(z) + fP_Q, \qquad (3)$$

where $P(z)$ is the curve in Figure 2. The result is the green curve shown in Figure 5. Whereas the microscopic $P(z)$ curves in each quantum, e.g., the one shown in Figure 2, are totally isolated from each other, $P_{macro}(z)$, the green line in Figure 5, extends into the adjacent quantum and is in fact identical with the blue curve in Figure 1A because both curves correspond to the same macroscopic current (the average position of the transitioning Cu atoms in the ACM scheme is the same as the position of hypothetical SCM Cu atoms).

The net message from Figures 2 and 5 is that the ACM scheme captures, on its own, the physics of both the macroscopic and microscopic length scales. The *P-ℰ* loops derived from PFM-measured *d-ℰ* loops reported in Ref. [3] reflect the fact that the microscopic polarization *dips* by -22 µC/cm$^2$ when the Cu atoms migrate across the vdW gaps from the +HP to the -HP state. In contrast, the *P-ℰ* loops predicted by the SCM approach in the previous section correspond to integrating the total macroscopic current, including the ionic-current component, and inevitably result in a *rise* of the polarization by +28 µC/cm$^2$.

**Acknowledgments:** We would like to thank Raffaele Resta and Petro Maksymovych for valuable discussions. Work at Vanderbilt was supported by the U.S. Department of Energy, Office of Science, Division of Materials Science and Engineering under Grant No. DE-FG02-09ER46554 and by the McMinn Endowment at Vanderbilt University. Work at ORNL was supported by the U.S. Department of Energy, Office of Science, Basic Energy Sciences, Materials Science and Engineering Division. Calculations were performed at the National Energy Research Scientific Computing Center, a DOE Office of Science User Facility supported by the Office of Science of the U.S. Department of Energy under Contract No. DE-AC02-05CH11231.